%% file: arxive_sub.tex
\documentclass[a4paper]{scrartcl}

\usepackage[utf8]{inputenc}
\usepackage[T1]{fontenc}
\usepackage[english]{babel}
\usepackage{amsmath}
\usepackage{amssymb}
\usepackage{bm}
\usepackage{dsfont}
\usepackage{enumitem}
\usepackage{authblk}
\usepackage{lipsum}
\usepackage{natbib}
\usepackage{url}
\usepackage[ruled, linesnumbered]{algorithm2e}
\usepackage{layouts}
\usepackage[font= footnotesize, labelfont=bf]{caption}
\usepackage{booktabs}
\usepackage{multirow}
\usepackage{dsfont}
\usepackage{comment}
\usepackage{graphicx}
\usepackage[dvipsnames]{xcolor}
\usepackage{soul}

\sethlcolor{BurntOrange}

\input{01_pkg_def.tex}

\title{Adaptive shrinkage of smooth functional effects towards a predefined functional subspace
}
\author{Paul Wiemann$^1$}
\author{Thomas Kneib$^1$}
\affil{$^1$G\"{o}ttingen University}
\date{\today}

\begin{document}
\maketitle

\begin{abstract}

In this paper, we propose a new horseshoe-type prior hierarchy for adaptively shrinking spline-based functional effects towards a predefined vector space of parametric functions.
Instead of shrinking each spline coefficient towards zero, we use an adapted horseshoe prior to control the deviation from the predefined vector space.
For this purpose, the modified horseshoe prior is set up with one scale parameter per spline and not one per coefficient.
The presented prior allows for a large number of basis functions to capture all kinds of functional effects while the estimated functional effect is prevented from a highly oscillating overfit.
We achieve this by integrating a smoothing penalty similar to the random walk prior commonly applied in Bayesian P-spline priors.
In a simulation study, we demonstrate the properties of the new prior specification and compare it to other approaches from the literature.
Furthermore, we showcase the applicability of the proposed method by estimating the energy consumption in Germany over the course of a day.
For inference, we rely on Markov chain Monte Carlo simulations combining Gibbs sampling for the spline coefficients with slice sampling for all scale parameters in the model.
\end{abstract}

\section{Introduction}\label{sec:introduction}

In this paper, we consider the non-parametric regression problem
\begin{align}\label{eg:model-eq-one-cv}
  y_i = f(x_i) + \varepsilon_i, \quad i = 1,\dotsc,n
\end{align}
with independent and identically distributed error terms ${\varepsilon_i \sim \ND(0, \sigma^2)}$, a smooth, nonlinear function $f$ and continuous covariate~$x_i$.
A common approach for solving the nonparametric regression problem is to transform it into a semi-parametric estimation problem by assuming that $f$ can be approximated via a linear combination of basis functions evaluated for the considered covariate \citep{ruppertSemiparametricRegression2003}.
In alternative to a regression spline, one might consider a parametric assumption on the dependence of the covariate, i.e., a linear or quadratic dependence, or a periodic dependence with a trigonometric polynomial.
To assess the correctness of the assumptions, goodness-of-fit measures like the widely applicable information criterion \citep[WAIC, ][]{watanabeAsymptoticEquivalenceBayes2010}, residual plots or posterior predictive checks can be employed \citep{gelmanBayesianDataAnalysis2014}.

Similar to \citet{shinFunctionalHorseshoePriors2020}, we present a methodology based on a shrinkage prior that represents a trade-off between the solution of a pre-defined parametric function and the solution of a flexible regression spline.
Consider a given, predefined subspace, e.g., the space of all linear or quadratic effects.
The prior we propose addresses the following three main objective:
\begin{itemize}
  \item For a weak signal or for a signal that can be represented well by a function from the pre-specified subspace, the estimated functional effect should be shrunken towards the considered subspace.
  \item When the signal is strong and it cannot be adequately represented by a function from the subspace, the shrinkage should be as small as possible.
  \item The prior should prevent the estimated function from oscillating too strongly and thus possibly overfitting the data. This is a known problem for unrestricted regression splines when many basis functions are used \citep{fahrmeirRegression2013}. However, we want the prior to be able to handle a high number of basis functions so that the space spanned by the basis functions can represent quickly changing functions.
\end{itemize}

The concept of shrinkage has successfully and widely been applied in many scientific fields since the foundational work by \citet{jamesEstimationQuadraticLoss1961}.
We mention here ridge regression \citep{hoerlRidgeRegressionBiased1970, marquardtRidgeRegressionPractice1975}, the LASSO penalty \citep{tibshiraniRegressionShrinkageSelection1996} and its Bayesian counterpart the Bayesian LASSO \citep{parkBayesianLasso2008} as examples for shrinkage of the regression coefficients in a linear model.
More recently, the concept of global-local shrinkage \citep{polsonShrinkGloballyAct2011} has come up.
Compared to fixpoint shrinkage, which shrinks every regression coefficient towards 0, global-local shrinkage shrinks the vector of all regression coefficients towards a sparse solution, which makes them in particular useful for high-dimensional but sparse estimation problems.
These global-local priors are usually equipped with global hyper-parameters that control the sparseness of the estimated vector.
The horseshoe prior \citep{carvalhoHorseshoeEstimatorSparse2010} is a prominent member of this class of shrinkage priors.
The horseshoe prior is especially known for its adaptive properties.
Due to the horseshoe-shaped prior on the shrinkage coefficient, it is capable of leaving strong signals almost untouched while applying strong shrinkage on weak signals.
\cite{polsonShrinkGloballyAct2011} draw a connection between global-local shrinkage and Bayesian model averaging that \citet{carvalhoHorseshoeEstimatorSparse2010} also highlight for their horseshoe prior.

Hence, the prior developed in this paper differs from the above mentioned priors, as we are not aiming at shrinking a sparse vector or individual regression coefficients towards zero.
Instead, we aim at shrinking the function estimated by a regression spline towards a predefined parametric subspace, while simultaneously controlling its curvature.
This distinguishes our approach from the goals usually pursued with the regularization of regression splines, which essentially focusses on limiting the curvature of the solution (see \citet{langBayesianPSplines2004}).

Recently, the work by \citet{shinFunctionalHorseshoePriors2020} has come close to and, by now, has inspired the method developed in this paper.
 However, their approach does not take the curvature of the estimated effect into account and, consequently, as we show in Section~\ref{sec:prior}, the estimated function becomes easily very wiggly or might not be capable to capture essential patterns in the data.
 Moreover, the method of effect selection introduced by \citet{scheiplSpikeandSlabPriorsFunction2012} and \citet{kleinBayesianEffectSelection2019} is not comparable with our approach.
 Firstly, instead of shrinkage, a spike and slab-based procedure is employed \citep{mitchellBayesianVariableSelection1988,ishwaranSpikeSlabVariable2005}.
 Secondly, the null effect arises from the kernel of the improper precision matrix in the prior on the spline coefficients and is therefore not an arbitrary parametric function of the covariate.

The remainder of this paper is organized as follows.
In Section~\ref{sec:prior}, we introduce the prior, discuss its extension for additive regression models (Section~\ref{sec:ex-to-star}), and investigate its shrinkage properties in Section~\ref{sec:null-space-and-tails}.
We describe the Bayesian inference scheme in Section~\ref{sec:inference} before demonstrating the validity of our approach with simulations in Section~\ref{sec:simulations}.
Here, we investigate the prior in a simple univariate case as well as in an additive model.
We conclude the empirical part of this paper in Section~\ref{sec:energy} with an application to energy consumption in Germany before we end with the discussion in Section~\ref{sec:discussion}.

\section{Definition of the smooth subspace shrinkage prior}\label{sec:prior}

In this section, we introduce the prior specification for estimating a smooth functional effect combined with shrinkage towards a predefined functional subspace.
Consider the non-parametric regression problem from Equation~\eqref{eg:model-eq-one-cv}.
As stated in the introduction, a common approach for the estimation of $f$ is to transform it into a semi-parametric estimation problem by assuming that $f$ can be approximated via a linear combination of $k$ pre-specified basis function denoted with $\Bfunc_j$. More precisely, we assume
\[
  f(x) = \sum_{j = 1}^{k} \Bfunc_j(x) \beta_j = \Bmat(x)'\betavec
\]
with $\Bmat(x) = (\Bfunc_1(x), \dotsc, \Bfunc_k(x))'$ and $\betavec = (\beta_1, \dotsc, \beta_k)'$.
Thus, a finite number of parameters needs to be estimated that have usually no valid interpretation by themselves.
In this work, we employ B-Splines of third order with equally spaced knots \citep{fahrmeirRegression2013}   .

\subsection{Common priors for splines}

The number of basis functions $k$ is highly important for the functional space spanned by the basis function.
A small number of basis functions may not be flexible enough to capture the pattern observed in the data.
Using many basis functions, however, may result in a highly oscillating function that overfits the data.
In the literature, different approaches can be found to overcome this problem.
One is to choose the number of basis function based on the number of available observations \citep[see][for details]{ruppertSemiparametricRegression2003}.

Let $\Zmat$ denote the $n \times k$ matrix of basis functions evaluated at the observed covariates $x_1, \dotsc, x_n$. When limiting the number of basis functions, Zellner's $g$-prior is a common choice for the prior on $\betavec$ \citep{zellnerAssessingPriorDistributions1986}
\[
  \betavec | g \sim \ND(\zerovec, g (\Zmat'\Zmat)^{-1})
\]
where the hyper-parameter $g>0$ adjusts the weight of the prior in relation to the observations.
Zellner's $g$-prior has the advantage of taking the (observed) correlation structure of the covariates into account.
Under this prior, the posterior mean of predictions given $g$ can be expressed as \[
  \EV(\hat\yvec | g, \yvec) = \left(1 - \frac{1}{1 + g}\right) \Zmat (\Zmat ' \Zmat)^{-1} \Zmat' \yvec,
\]
which shows that $g$ can be interpreted as a shrinkage parameter with shrinkage towards the null effect.
In this approach, the regularization of the spline is achieved by limiting the number of basis functions depending on the number of observations.
Thus the functional pattern that can be represented with the spline is dependent on the number of observations.
This is not a desirable property.

To circumvent this issue, a different approach is based on regularization of the curvature of estimated function or, to word it differently, shrinkage towards a function with less curvature.
Here, a moderately large number of basic functions is used in conjunction with a prior which ensures that adjacent regression coefficients are not too different, and thus produces a certain smoothness of the functional estimate.

The Bayesian P-spline approach by \cite{langBayesianPSplines2004} achieves this by placing a Gaussian random walk prior on the spline coefficients.
In the case of a first order random walk $\beta_j = \beta_{j-1} + u_j$ for $1 < j \le k $, with $u_j \sim \ND(0, \tau^2)$, the variance of the increments acts as a smoothness parameter and the prior shrinks towards a constant functional effect.
The second order random walk $\beta_j = 2 \beta_{j-1} - \beta_{j-2} + u_j$, for $2 < j \leq k$, acts as a linear extrapolation with a Gaussian error term $u_j \sim \ND(0, \tau^2)$.
Again, $\tau^2$ can be interpreted as a smoothing parameter and the prior shrinks towards a linear function.
The overall level is not penalized by the Bayesian P-spline prior since it places a flat prior on $\beta_1$ and for the second order random walk the flat prior extends to $\beta_2$ and thus the slope is not regularized.
With the appropriate rank deficient penalty matrix $\Kmat$, this prior can equivalently expressed as
\[
  \betavec \sim \NDprec(\zerovec, \tau^{-2} \Kmat)
\]
where $\NDprec$ is the degenerated normal distribution with precision matrix $\tau^{-2}\Kmat$ (see \cite{langBayesianPSplines2004} and \cite{rueGaussianMarkovRandom2005} for details).

For the construction of our smooth subspace shrinkage, we use a combination of both approaches: we use the smoothing properties of the Bayesian P-spline and combine it with an extended version of Zellner's $g$-prior - very similar to the functional horseshoe prior \citep{shinFunctionalHorseshoePriors2020} - to shrink towards a predefined functional subspace.

\subsection{Construction of the smooth subspace shrinkage prior}

In comparison to the Bayesian P-spline, we do not only want to shrink towards less curvature, but additionally to a parametric functional subspace of the covariate.
To give an example, this could be a quadratic effect, as well as a periodic effect.
Nevertheless, the key feature of the Bayesian P-spline should be maintained, namely preventing the functional estimate from exhibiting highly oscillating behavior and thus overfitting the data when a high number of basis functions is used.
The prior however should be able to adapt to quickly changing functions when the data is suggesting this, so we waive the option of using only a small number of basis functions.

We achieve our goal by using the following prior hierarchy.
Let $\Smat$ be a matrix whose columns span the space $\mathcal{N}$ towards we want to shrink.
To give examples, use $\Smat = (\onevec, \xvec)$ for the null space comprised by the linear effect and $\Smat = (\onevec, \xvec, \xvec^2)$ when quadratic effects should comprise the null space.
Hereby, operations on the covariate vector are carried out element-wise.
Based on $\Smat$, we define the $n \times n$ matrices $\Pmat_0 = \Smat(\Smat'\Smat)^{-1}\Smat$ and $\Pmat_1 = \Imat - \Pmat_0$ where $\Imat$ denotes the identity matrix of appropriate size.
$\Pmat_0$ and $\Pmat_1$ are both projection matrices with their images being orthogonal complements of each other.
 Additionally, the union of their images is equal to the space spanned by the columns of $\Zmat$.

By construction, $\mathcal{N}$  is the image of $\Pmat_0$ and the kernel of $\Pmat_1$.
We can now decompose the image of the element-wise application of the function $f(\xvec) = (f(x_1), \dotsc, f(x_n))'$ into the part within and not within the null space, i.e.,
\[
  f(\xvec) = (\Pmat_0 + \Imat - \Pmat_0)f(\xvec) = \underbrace{\Pmat_0 f(\xvec)}_{\in \mathcal{N}} + \underbrace{\Pmat_1 f(\xvec)}_{\in\bar{\mathcal{N}}}
\]
where $\bar{\mathcal{N}}$ denotes the orthogonal complement of $\mathcal{N}$.
Therefore, $f^{\bar{\mathcal{N}}}(\xvec) = \Pmat_1 f(\xvec)$ is the part of the functional effect that is not within the null space and consequently should be controlled by the shrinkage prior. To achieve this, we propose to place the improper prior
\begin{align*}
  \betavec | \lambda, \tau^2, \sigma^2 \sim \NDprec(\zerovec, \Qmat) \text{ with } \Qmat = \sigma^{-2} \lambda^{-2} \Zmat'\Pmat_1\Zmat + \tau^{-2}\Kmat
\end{align*}
on $\betavec$ where $\Kmat$ denotes the appropriate second-order walk penalty matrix.
Therefore,
\[
  \betavec'\Qmat\betavec = \frac{f^{\bar{\mathcal{N}}}(\xvec)'f^{\bar{\mathcal{N}}}(\xvec)}{\lambda^2\sigma^2} + \frac{\betavec'\Kmat\betavec}{\tau^{2}};
\]
i.e., the argument to the exponential function in the degenerated normal density of the prior on $\betavec$
measures the sum of the quadratic deviation from the functional null space (weighted with $\lambda^{-2}\sigma^{-2}$) in the first term and the sum of the quadratic second order differences of the spline coefficients (weighted with $\tau^{-2}$) in the second term.
We deal with the priors on the scale parameters later in Equation~\eqref{eq:prior-scale}.

The posterior of $\betavec$ conditioned on $\sigma^2, \lambda$ and $\tau^2$ is then normally distributed with mean $\betavec^\star$ and precision matrix $\Qmat^\star$ given as follows:
\begin{align} \label{eq:beta-posterior}
  \Qmat^\star &= \sigma^{-2}\Zmat'\Zmat + \Qmat &\betavec^\star &= \sigma^{-2}(\Qmat^\star)^{-1}\Zmat' \yvec.
\end{align}
When defining 
\begin{align}
  \kappa = \frac{1}{1 + \lambda^2}\label{eq:def-kappa}
\end{align}
as in the horseshoe prior and considering the expected functional effect conditioned on no penalization of the second order derivative
\begin{align}\label{eq:posterior-expectation}
  \lim_{\tau^2 \to \infty} E(\Zmat\betavec |  \yvec, \lambda, \tau^2) = ((1 - \kappa) \Pmat_\Zmat + \kappa \Pmat_0)\yvec
\end{align}
where $\Pmat_\Zmat = \Zmat (\Zmat' \Zmat)^{-1} \Zmat'$, the role of $\kappa$ as the shrinking factor becomes apparent as the expectation is a weighted average of the parametric least squares solution (i.e., $\Pmat_0\yvec$)  and the unpenalized spline estimate (i.e., $\Pmat_Z \yvec$).
We reach the extreme of no shrinkage and the reassembling of the spline solution for $\kappa \to 0$. Full shrinkage is obtained for $\kappa \to 1$, which equals the parametric least square solution.

We now discuss the necessity of the second term in the precision matrix for the prior on $\betavec$.
From Equation~\eqref{eq:posterior-expectation}, it is evident that the spline solution passes on its highly oscillating behavior to the weighted average if there is no, or only little, shrinkage.
No shrinkage or slight shrinkage is especially desirable in the situation in which the defined null space does not match the pattern in the data, and thus the model is supposed to adapt to the data.
But even in the case of strong but not complete shrinkage, the highly oscillating behavior is included in the estimated function albeit with smaller amplitude.

\begin{figure}
  \includegraphics[width=\textwidth]{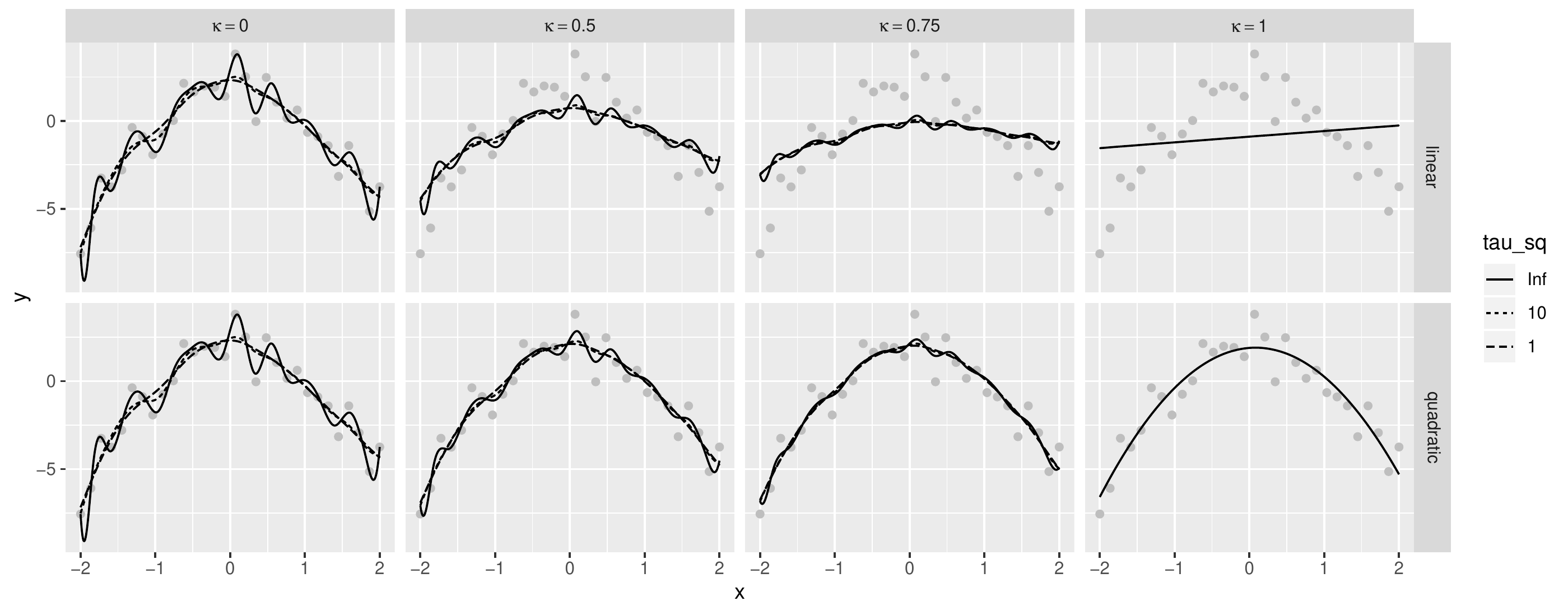}
  \caption{The plots demonstrate the effect of the shrinkage coefficient~$\kappa$ and the smoothness variance~$\tau^2$ on the posterior mean as given in Equation~\eqref{eq:beta-posterior} for a data example (scatterplot) generated from a quadratic function.
  The lines represent the posterior mean. Without any smoothing (i.e., $\tau^2 \to \infty$) the spline is highly oscillating and overfitting the data when no, or only slight, shrinkage is present. In this example, $\sigma^2 = 1$ and the spline is based on 20~basis functions.
  \label{fig:oscillating}}
\end{figure}

We illustrate this with an example.
Consider the solid line in Figure~\ref{fig:oscillating}.
Without any penalization of the second order derivative (i.e., $\tau^2 \to \infty$) the expected value of the prediction tends to overfit the data.
Only in the case of the correct specified null space, we see a fitting and not oscillating functional estimate (second row, fourth column). In addition, the figure demonstrates that integrating the second order random walk prior on $\betavec$  helps to prevent this.
Even when no shrinkage is present (i.e., $\kappa \to 0$), the posterior mean exhibits less curvature and fits the underlying pattern better.

 We place half Cauchy and inverse gamma priors on the scale parameters in the prior on $\betavec$
\begin{align}
  \lambda &\sim \CD_+(0, \tilde\xi),  &\xi &\sim \CD_+(0, \xi_0), &\tau &\sim \CD_+(0, \nu) \quad \text{and} &\sigma^2 \sim \InvGD(a_0, b_0)\label{eq:prior-scale}
\end{align}
where $\CD_+(a, b)$ denotes the Cauchy distribution with mean parameter $a$ and scale parameter $b$ truncated to the positive real numbers and $\InvGD(c, d)$ denotes the inverse gamma distribution with scale parameter $c$ and shape parameter $d$.

A probability statement (Equation~\eqref{eq:prob-nu}) is used to determine the prior on the variance of the random walk, namely $\nu > 0$.
Over the domain of the covariates, the second order derivative should be, in absolute terms, smaller than the prespecified threshold $c$ with probability $1 - \alpha$, i.e.,
\begin{align}\label{eq:prob-nu}
  \lim_{\lambda \to \infty} \Pr\left(\left.\max_{x \in \mathcal{D}}\left|f''(x)\right| < c \right| \lambda \right) = 1 - \alpha.
\end{align}

Furthermore,  $\xi_0, a_0, b_0 > 0$ must be specified by the user and $\tilde \xi$ is a function of $\xi$.
The prior on $\lambda$ is an adaptation of the hierarchy considered in the horseshoe prior \citep{carvalhoHorseshoeEstimatorSparse2010}.
We scale $\xi$ such that the marginal variance of the spline coefficients is not dependent on the design matrix of the spline $\Zmat$ nor the defined null space and thus the projection matrix $\Pmat_1$.
Following the idea of \cite{sorbyeScalingIntrinsicGaussian2014}, we define
\[
  \tilde\xi = \xi \sigma^{-1}_{\text{ref}(\Zmat'\Pmat_1\Zmat)}
\]
where $\sigma^{-1}_{\text{ref}(\cdot)}$ represents average marginal standard deviation induced by the precision matrix.
For that we calculate the standard deviation as the geometric mean
\[
  \sigma_{\text{ref}(\Qmat)} = \exp\left(\frac{1}{n}\sum_{i=1}^{n}\frac{1}{2}\log(\bm{\Sigma}_{ii}^\star)\right)
\]
with $\bm{\Sigma}^{\star}$ as the generalized inverse of $\Qmat$.

To have more control over the shrinkage properties, especially the strength of a signal, $\xi$ may be fixed.
This might be in particular well suited in situations in which only one functional effect should be controlled by the proposed prior and, therefore, no global-local shrinkage is required. The extension to multiple additive effects is described in  Section~\ref{sec:ex-to-star} and an assessment of the shrinkage properties is made in Section~\ref{sec:null-space-and-tails}.

\subsection{Extension to multiple additive functional effects}\label{sec:ex-to-star}

We describe briefly how the proposed prior hierarchy can be extended to cover multiple functional effects in an additive regression design \citep{hastieGeneralizedAdditiveModels1986}.
Suppose $L$ smooth effects should be modeled such that the response for $i=1,\dotsc,n$ observations is given by
\[
  y_i = \beta_0 + f_1(x_{i1}) + f_2(x_{i2}) + \dotsc + f_L(x_{iL}) + \varepsilon_i
\]
where the distributional assumption from above applies to $\varepsilon_i$.
In addition to the assumption that the smooth functions can be represented by a spline, we assume for identification purposes that the smooth functions are centered; i.e., the sum over the coefficients of each spline equals 0.
This assumption is justified because it causes no general restriction that cannot be nullified by centering $\yvec$ or by the inclusion of an intercept, i.e., $\beta_0$.

As above, we approximate the smooth effects with linear combinations of basis functions resulting in the design matrices $\Zmat_1, \dotsc, \Zmat_L$ with the corresponding vectors of coefficients $\betavec_{1}, \dotsc, \betavec_{L}$.
For $l = 1, \dotsc, L$, the prior hierarchy on the coefficient vectors is given by
\begin{align*}
  &\betavec_l | \lambda_l, \tau_l, \sigma^2 \sim \NDprec(\zerovec, \Qmat_l) &\text{ with } \Qmat_l = \sigma^{-2} \lambda_l^{-2} \Zmat_l'\Pmat^{(l)}_1\Zmat_l + \tau^{-2}\Kmat_l\\
  & &\text{ and the linear restriction }\onevec'\betavec_l = 0\\
  &\lambda_l | \xi \sim \CD_+(0, \tilde\xi_l) & \text{ with } \tilde\xi_l = \xi \sigma^{-1}_{\text{ref}(\Zmat_l'\Pmat^{(l)}_1\Zmat_l)}\\
  &\xi \sim \CD_+(0, \xi_0) &\\
  &\tau_l | \nu_l \sim \CD_+(0, \nu_l) &\hspace{-1cm}\text{with $\nu_l$ s.t.} \lim_{\lambda_l \to \infty} \Pr\left(\left.\max_{x \in \mathcal{D}_l}\left|f_l''(x)\right| < c_l \right| \lambda_l \right) = 1 - \alpha_l
\end{align*}
where
\begin{itemize}
  \item the prior on $\betavec_l$ includes now the linear restriction $\bm{1}'\betavec_l = 0$ for identifiability,
  \item $\Kmat_l$ denotes the appropriate second order walk penalty matrix,
  \item $\Pmat^{(l)}_1$ denotes the projection matrix with kernel equal to the chosen null space for the $l$-th effect,
  \item and $\alpha_l$ and $c_l$ are chosen as above to control the wigglyness of $f_l$.
\end{itemize}
In addition, we place the same prior as before on $\sigma^2$ and use a non-informative prior for the intercept $\beta_0$.
In this prior hierarchy and similar to the horseshoe prior, the prior distributions on $\lambda_l$ are connected via $\xi$ which acts as a global shrinkage parameter while $\lambda_l$ controls the local shrinkage.

\subsection{Behavior in the null space and tail behavior}\label{sec:null-space-and-tails}

In this section, we study the shrinkage properties of the proposed prior.
The literature suggests \citep{scheiplSpikeandSlabPriorsFunction2012,kleinBayesianEffectSelection2019} to examine the marginal distribution of the spline coefficients, i.e., $p(\betavec_l)$ for the $l$-th functional effect.
For simplicity, we make the same assumption as \cite{carvalhoHorseshoeEstimatorSparse2010} regarding the variance of the error term, namely $\sigma^2 = 1$.
The error variance can be easily estimated from the data and acts in the proposed prior hierarchy just as a scaling component.
In addition, we anticipate that the prior on the second order random walk, specifically on $\tau_l$, has been chosen such that it does not affect the shrinkage properties.
We consider this assumption to be valid, as we believe that the user selects the limit for the second derivative to prevent overfitting, but not to affect the shrinkage properties.
Therefore, we consider in the following the limit $\tau_l \to \infty$.
For the sake of compactness, we refrain from stating the conditioning on $\tau_l$ and $\sigma^2$ for the rest of this section.

The proposed prior features local-global shrinkage properties with the local shrinkage parameter $\lambda_l$ and the global shrinkage parameter $\xi$.
Since i) the main local shrinkage properties are unaffected by the global shrinkage parameter's value and ii) the effect on the marginal distribution of $\betavec_l$ without a fixing global shrinkage parameter depends on the specific predictor structure, we condition in the following on $\xi$ and thus $\tilde \xi_l$.

Note that the results derived in this section do not depend on a specific value of $\xi$. For a better readability we drop in the following the index $l$.
The implied marginal distribution of $\betavec | \tilde\xi$ can now be derived as
\begin{align*}
  p(\betavec | \tilde\xi) &= \int_0^\infty p(\betavec | \lambda)p(\lambda | \tilde\xi^2) d\lambda \\
    &= (2\pi)^{-\rank(\Fmat)/2} (|\Fmat|^*)^{1/2} \int_0^\infty \lambda^{-\rank(\Fmat)}\exp\left(-\frac{1}{2\lambda^2}\betavec'\Fmat\betavec\right) \frac{2}{\pi\tilde\xi}\left(1 + \frac{\lambda^2}{\tilde\xi^2}\right)^{-1} d\lambda\\
    &= \underbrace{\frac{2}{\pi\tilde\xi}(2\pi)^{-\rank(\Fmat)/2} (|\Fmat|^*)^{1/2}}_{=:c_0} \int_0^\infty \frac{\exp(-\frac{1}{2\lambda^2}\betavec'\Fmat\betavec)}{\lambda^{\rank(\Fmat)}\left(1 + \frac{\lambda^2}{\tilde\xi^2}\right)}d\lambda.
\end{align*}
where $\Fmat = \Zmat'\Pmat_1\Zmat$.
No analytical solution exists for this integral, though.
We define the constant scalar in front of the integration symbol as $c_0$.

\begin{figure}[hbt]
  \begin{center}
    \includegraphics[width=\textwidth]{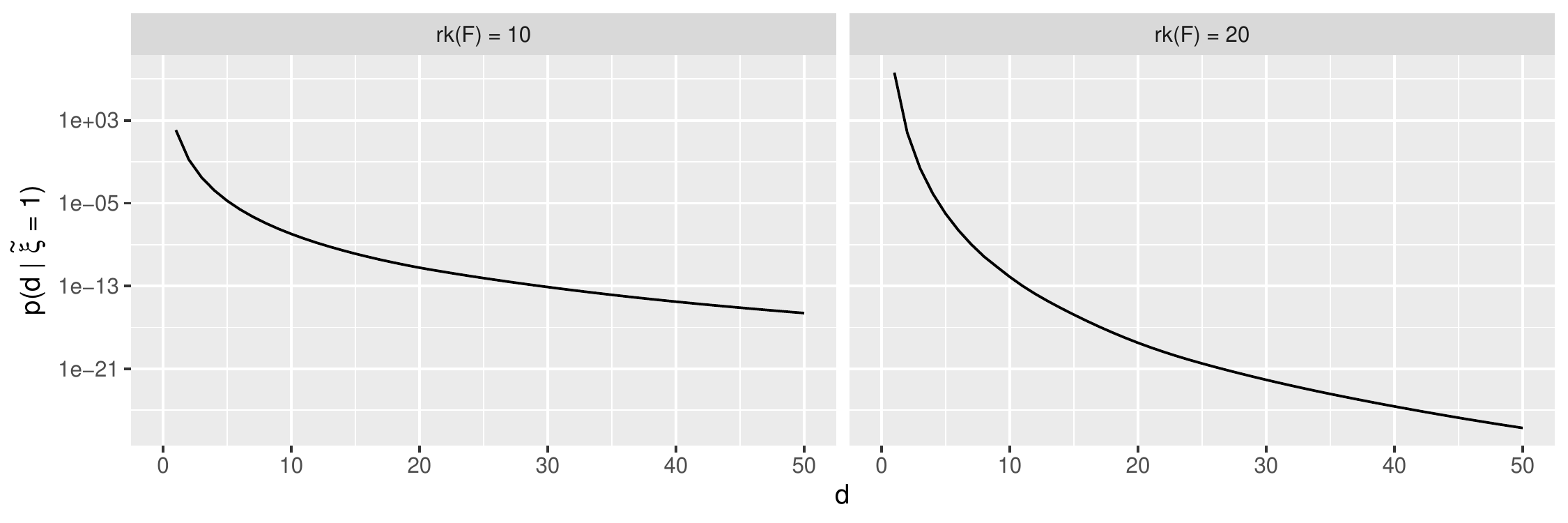}
    \caption{Plots of the marginal density of the distance of spline coefficients to the null space conditioned on $\tilde\xi$. The left plot is based on $\rank(\Fmat) = 10$ and the right plot on $\rank(\Fmat) = 20$.}
    \label{fig:marginal-distance}
  \end{center}
\end{figure}

In the assessment of shrinkage properties, we are mainly interested in the tail behavior and the behavior in the origin with respect to the distance of the functional effect from the null space $\mathcal{N}$.
To assess this, we define this distance as $d = ||\Pmat_1 \Zmat \betavec||$. The implied marginal density on $d$ is then given as
\begin{align*}
  p(d | \tilde\xi) &= c_0 \int_0^\infty \frac{\exp(-\frac{d^2}{2\lambda^2})}{\lambda^{\rank(\Fmat)}\left(1 + \frac{\lambda^2}{\tilde\xi^2}\right)}d\lambda
\end{align*}
with no closed form solution available so that we approximate it numerically for a visual impression shown in Figure~\ref{fig:marginal-distance}.

We investigate more closely, whether the spike in the origin is infinite. An infinite spike in zero exhibits particular advantageous shrinkage properties as it implies a strong penalization and thus strong shrinkage towards the origin for small effects \citep{kleinBayesianEffectSelection2019}.
Translated to our hierarchy, this implies that effects close to the null space receive strong shrinkage towards the null space.
Using the last equation, we obtain
\begin{align*}
  p(d = 0| \tilde\xi) &= c_0 \int_0^\infty \frac{1}{\lambda^{\rank(\Fmat)}\left(1 + \frac{\lambda^2}{\tilde\xi^2}\right)}d\lambda \\
    &=
      c_0\int_0^1 \underbrace{\frac{1}{\lambda^{\rank(\Fmat)}\left(1 + \frac{\lambda^2}{\tilde\xi^2}\right)}}_{ =: w(\lambda)} d\lambda +
      \underbrace{c_0 \int_1^\infty \frac{1}{\lambda^{\rank(\Fmat)}\left(1 + \frac{\lambda^2}{\tilde\xi^2}\right)}d\lambda}_{\geq 0}
\end{align*}
for which we can find a bound from below for the integrand $w(\lambda)$ as
\begin{align*}
  w(\lambda) &= \frac{1}{\lambda^{\rank(\Fmat)} + \frac{\lambda^{\rank(\Fmat) + 2}}{\tilde\xi^2}} = \frac{\tilde\xi^2}{\tilde\xi^2\lambda^{\rank(\Fmat)} + \lambda^{\rank(\Fmat) + 2}}\\
    &\stackrel{0 < \lambda < 1}{\geq} \frac{\tilde\xi^2}{\tilde\xi^2\lambda^{\rank(\Fmat)} + \lambda^{\rank(\Fmat)}}
\end{align*}
and thus
\begin{align*}
  \int_0^1 w(\lambda) d\lambda &\geq \int_0^1 \frac{\tilde\xi^2}{\tilde\xi^2\lambda^{\rank(\Fmat)} + \lambda^{\rank(\Fmat)}}\\
    &\stackrel{\rank(\Fmat) > 1}{=} \left[\frac{\tilde\xi^2\lambda^{1 - \rank(\Fmat)}}{(1 - \rank(\Fmat))(\tilde\xi^2 + 1)}\right]_0^1 = \infty
\end{align*}
The marginal has indeed an infinite spike in the origin w.r.t.\ $d$ and consequently the marginal $p(\betavec)$ has an infinite spike if $\Zmat\betavec$ is in the null space.

\begin{figure}[bt]
  \begin{center}
    \includegraphics[width=\textwidth]{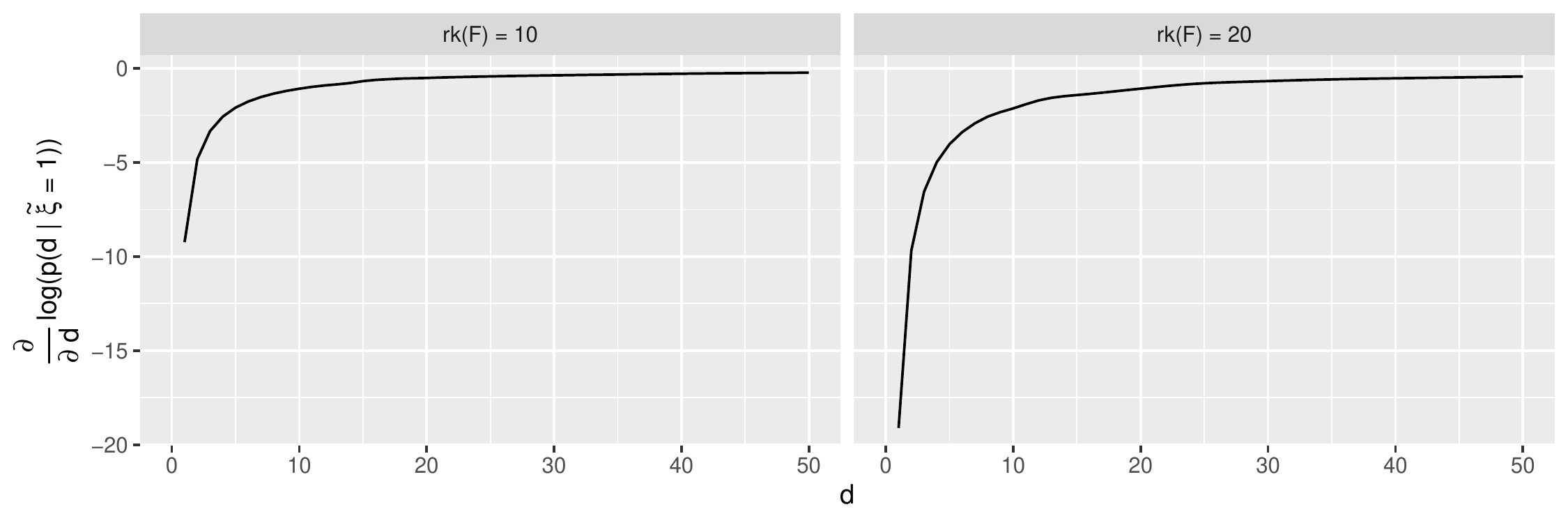}
    \caption{Plots of the marginal score function of the distance of spline coefficients to the null space conditioned on $\tilde\xi$. The left plot is based on $\rank(\Fmat) = 10$ and the right plot on $\rank(\Fmat) = 20$.}
    \label{fig:marginal-distance-score}
  \end{center}
\end{figure}

Besides studying the marginal prior in the origin, \cite{scheiplSpikeandSlabPriorsFunction2012} suggest to explore the tail behavior by focusing on the first derivative of the log marginal, i.e., the score function and its properties in the limit.
\cite{kleinBayesianEffectSelection2019} state that the prior has heavy tails, if the score function is redescending; namely, its value approaches zero in the infinite limit of its argument.
Heavy tails in the marginal prior are desirable since they imply Bayesian robustness of the estimates.
Unfortunately, the limiting behavior of the score function
\begin{align*}
  \frac{\partial}{\partial d} \log(p(d | \tilde\xi^2))
  &= \frac{\frac{\partial}{\partial d} p(d | \tilde\xi^2)}{p(d | \tilde\xi^2)}\\
  &= \frac
    {-d c_0 \int_0^\infty \frac{\exp(-\frac{d^2}{2\lambda^2})}{\lambda^{\rank(\Fmat) + 2}\left(1 + \frac{\lambda^2}{\tilde\xi^2}\right)}d\lambda}
    {c_0 \int_0^\infty \frac{\exp\left(-\frac{d^2}{2\lambda^2}\right)}{\lambda^{\rank(\Fmat)}\left(1 + \frac{\lambda^2}{\tilde\xi^2}\right)}d\lambda} \leq 0
\end{align*}
is not analytically accessible so that we must rely on a numerical approximation. Figure~\ref{fig:marginal-distance-score} displays plots of the score function.
From the plots we conclude that the prior constitutes a redescending score function and thus has heavy tails.
These findings translate to the functional space which consequently means that functional estimates in the tails of the prior are robust.

\section{Bayesian inference}\label{sec:inference}

We base our Bayesian estimation procedure on Markov chain Monte Carlo (MCMC) simulations and employ a within-Gibbs updating scheme.
Within each MCMC iteration, we initially update the intercept $\beta_0$ with a Gibbs update and then update the spline coefficients of each smooth function with a draw from the full conditionals.

Suppose the $l$-th component of the predictor should be updated.
Let $\tilde \yvec_{-l}$ denote the working observations, that is $\tilde \yvec_{-l} = \yvec - \etavec_{-l}$ where $\etavec_{-l}$ denotes the predictor with the effect of the component under consideration removed.

With a non-informative prior on the intercept, the full conditional distribution of the intercept is then given by $\beta_0 |\, \cdot \sim \ND\left(\bar{\tilde\yvec}_{-0}, \sigma^2 / n\right)$ where $\bar{\tilde\yvec}_{-0}$ is the mean of the working observations.

For the spline coefficients of the $l$-th smooth effect, the full conditional follows from Equation~\eqref{eq:beta-posterior}, and is given by
\[
  \betavec_l |\, \cdot \sim \ND\left(
    \betavec_l^\star,
    (\Qmat_l^\star)^{-1}
  \right)
  \text{ where }
  \Qmat_l^\star = \sigma^{-2}\Zmat_l'\Zmat_l + \Qmat_l
  \text{ and }
  \betavec_l^\star = (\Qmat_l^\star)^{-1}\Zmat_l' \tilde\yvec_{-l}.
\]
To ensure identifiability, we sample under the linear constrain $\onevec'\betavec_l = 0$ as described in \citet[Algorithm~2.6]{rueGaussianMarkovRandom2005}.

Finally, the scale parameters of the model are updated using slice updates \citep{nealSliceSampling2003}.
Hereby, we choose to update first  $\sigma^2$, then the scale parameters of each spline, namely $\lambda_i$ and $\tau_i$, and lastly the global shrinkage parameter $\xi$.
All updates of the scale parameters are performed in log-space.
As proposed by \cite{makalicSimpleSamplerHorseshoe2016}, the global shrinkage parameter could also be updated using the double inverse gamma representation.
However, we could not observe improvements of the MCMC estimation.
Due to that, we stick to the slice sampler for the sake of consistency.

\section{Simulations}\label{sec:simulations}

We assess the validity of our approach in three simulation studies.
In the first two studies, we employ a simple additive model featuring one covariate. 
We explore the behavior of the presented prior considering different null spaces and its reaction to different signal to noise ratios (SNRs).
In the third simulation study, we analyze the proposed prior within the additive model and compare the results to the popular Bayesian P-splines approach.

Notably, the work in \citet{shinFunctionalHorseshoePriors2020} has partly close resemblance to the method developed in this paper.
Nonetheless, we refrain from an empirical comparison to this approach, since the functional horseshoe prior lacks a smoothing component.
This is not in line with our goal of producing flexible and non-wiggly functional estimates.

\subsection{Simple additive model design}\label{sec:sim-one-cv}

In the first two scenarios, we consider a simple regression setting with only one smooth effect and 100~observations.
A quadratic effect is featured in scenario~I, while a more complex function is used in scenario~II. In particular, the smooth functions are defined as
\begin{align*}
  f(x) &= \frac{1 + 1.5x^2}{20} &\text{ (scenario I) and }\\f(x) &= \frac{1 + 10\sin(x) + x + 0.64x^2}{20} &\text{ (scenario II)}.
\end{align*}
The covariate values $x$ are equally spaced within the interval $[-2\pi, 2\pi]$.
In both scenarios, we add an independently and normally distributed error term with zero mean and variance equal to $\sigma^2$.

We fit models comprising the proposed prior with null spaces spanned by $[\bm{1}, \xvec]$ and $[\bm{1}, \xvec, \xvec^2]$ in scenario~I and add in scenario~II the null spaces spanned by $[\bm{1}, \sin(\xvec)]$ and $[\bm{1}, \xvec, \xvec^2, \sin(\xvec)]$.
The operations on the covariates are defined element-wise and the last null space is referred to as complex in the following.
To feature two different SNRs, we choose the standard derivation of the error term $\sigma$ equal to $0.75$ or equal to $2.5$.
Each scenario is replicated 100~times.

For the estimation, we employ a cubic spline with 20~inner knots and we determine the parameter $\nu$ in the prior on $\tau$ based on $\alpha = 0.05$ and the cutoff $c$, where $c$ is determined as follows.
Let $c_p$ be the maximum absolute value of the second order derivative of the parametric solution within the null space. Then the cutoff is set to $c = 10 \max(c_p, 0.1)$.
Furthermore, $\xi_0$ is set to 1.
Inference is based on 10.000~MCMC iterations of which the first half is considered warm-up.

\begin{figure}[bht]
  \begin{center}
    \includegraphics[width=\textwidth]{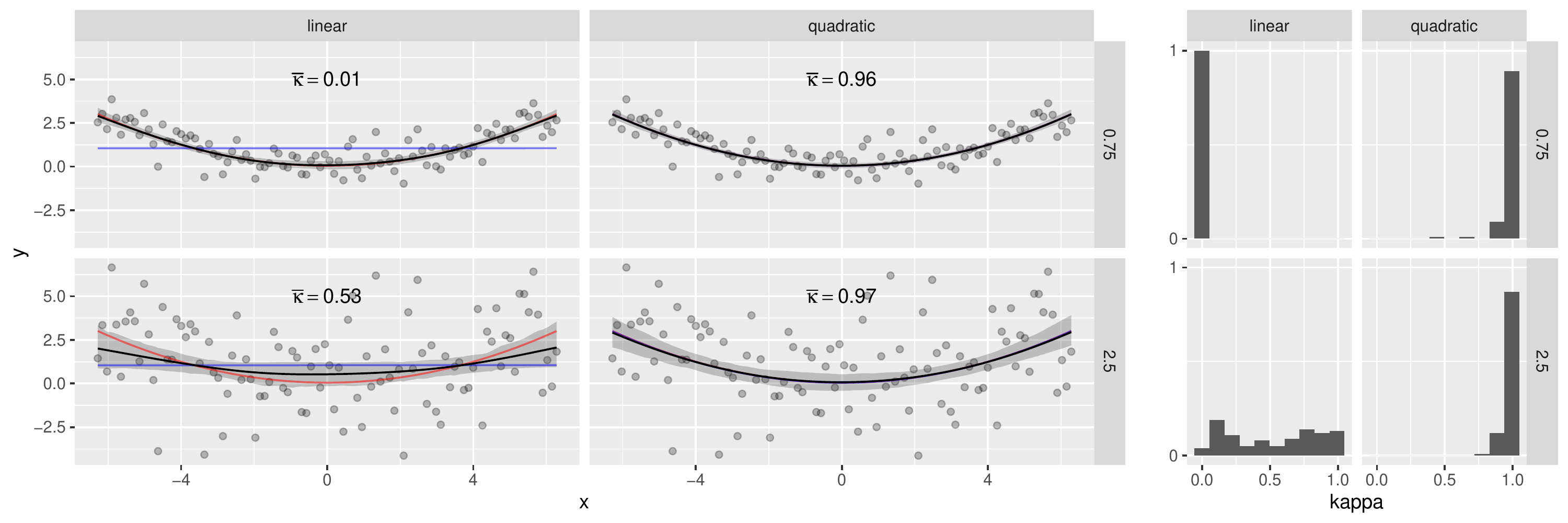}
    \caption{Plots concerning simulation scenario~I. The left plot shows data of one iteration, the data generating function (red line), the average of the posterior mean using a parametric model (blue line), and the average of the posterior means together with 90\% quantiles when employing the proposed model. The right plot shows histograms of the posterior mean of the shrinkage parameter $\kappa$ (see Equation~\eqref{eq:def-kappa}).
    The columns within the plots distinguish between different types of null spaces which are also used for the parametric estimates.
    Each row within the plots displays a different SNR scenario.}
    \label{fig:sim1}
  \end{center}
\end{figure}

\begin{figure}[bt]
  \begin{center}
    \includegraphics[width=\textwidth]{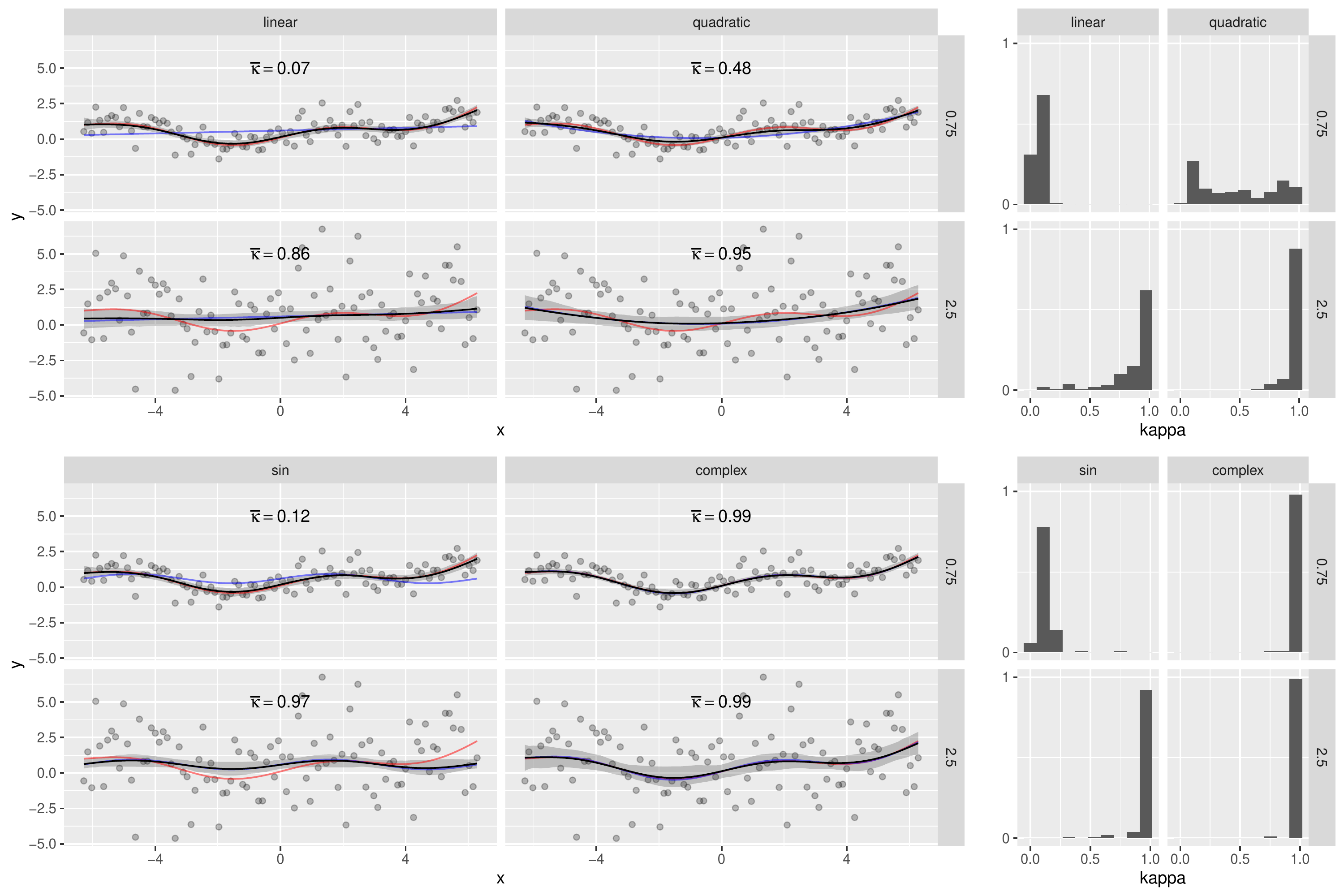}
    \caption{Plots analog to Figure~\ref{fig:sim1} but concerning simulation scenario~II. The second row of plots comprises the additional null spaces.
    \label{fig:sim2}
    }
  \end{center}
\end{figure}

The main results are summarized in Figure~\ref{fig:sim1} and Figure~\ref{fig:sim2} for scenario~I and scenario~II, respectively.
In scenario I, we can see that the prior adopts to the data generating function regardless of the specified null space, when the signal is strong enough.
With more noise, the prior still shrinks to the parametric function when the correct null space is specified.
Concerning the linear null space, the prior performs worse and does not show a consistent behavior when more noise is present.

Now focusing on scenario~II, we can deduce that the prior is mostly able to decide between signal and no signal as most values are either close to 0 or 1.
Furthermore, we can observe that in the high noise scenario the prior forces the estimate to be very close to the parametric solution and thus to be in the null space.
This is especially true in both SNR scenarios for the \emph{complex} null space which includes the data generating function.
In the low-noise scenario, the estimate is mainly able to capture the true function with some difficulties with the \emph{quadratic} null space.

\subsection{Additive model design}\label{sec:sim-star}

In the final simulation study, scenario~III, we use the proposed prior within an additive predictor that includes the effect of multiple covariates.
Its performance is compared to Bayesian P-splines \citep{langBayesianPSplines2004}.
To do so, we generate data for ${n=100}$~observations according to
\[
  y_i = f_1(x_{i1}) + f_2(x_{i2}) + f_3(x_{i3}) + f_4(x_{i4}) + \varepsilon_i \qquad\text{for } i=1,\dotsc, n
\]
with the covariate information $x_{ij}$ independently and uniformly distributed on the interval $[-1, 1]$.
Furthermore, $\varepsilon_i$ is independently normal distributed with expectation zero and variance $\sigma^2$ and
$f_1, \dotsc, f_4$ denote four smooth functions, in particular
\begin{align*}
  f_1(x) &= x
  &f_2(x) &= 2x^2 - 3/2\\
  f_3(x) &= \sin(x\pi)
  &f_4(x) &= \frac{2\exp(x)}{\exp(1) - \exp(-1)} - 1
\end{align*}
where the integral from -1 to 1 is equal to 0 for all functions and they are scaled such that the Lebesgue measure of image of $[-1, 1]$ under the function is equal to 2; that is, $\sup(\{f(x) | x \in [-1, 1]\}) - \inf(\{f(x) | x \in [-1, 1]\}) = 2$ for a continuous function $f$ on $[-1, 1]$.

We fit two models to the data.
In the first model, each smooth function is represented by Bayesian P-splines with the default inverse gamma prior ($a = 0.001, b = 0.001$) on the variance parameter of spline coefficients.
In the second model, the smooth functions are estimated employing the introduced prior.
In each model, we use the same number of inner knots (i.e.~20) and the same knot positions.
For the shrinkage prior, specify the correct null space for the first three functions, i.e., $[1, \xvec_1]$, $[1, \xvec_2, \xvec_2^2]$, and $[1, \sin(\xvec_3 \pi), \cos(\xvec_3 \pi)]$.
As before, the operations on the vector are interpreted element-wise.
For the fourth smooth effect, the null space comprises, in contrast to the true effect, only of constant effects.
We set $\nu_l= 0.1$ for $l=1,\dotsc,4$ and $\xi_0 = 1$.
The scenario is replicated 100~times and we use 15.000~MCMC iterations in the estimation procedure of which the first half is considered warm-up.

\begin{figure}[tb]
  \begin{center}
    \includegraphics[width=\textwidth]{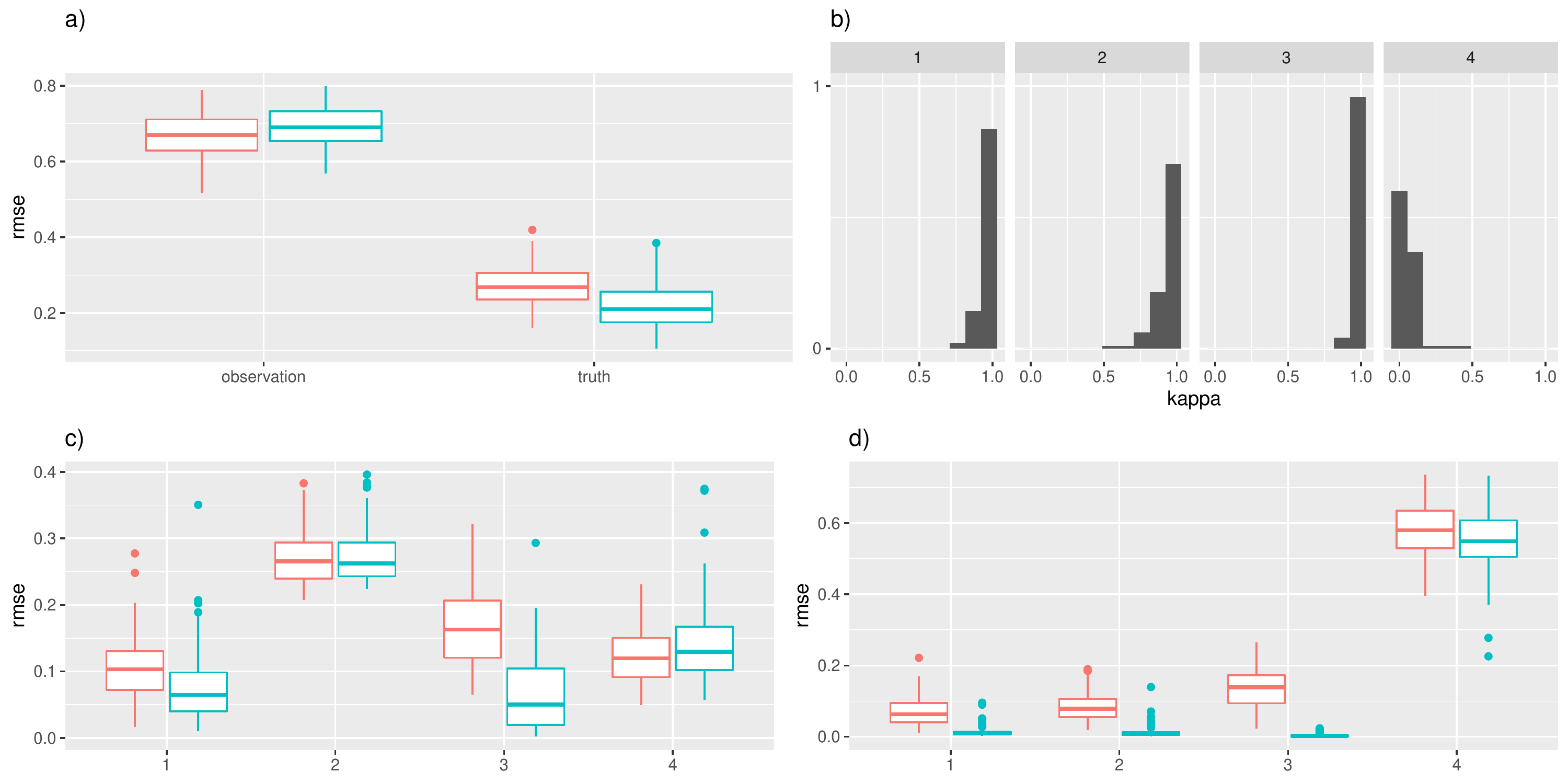}
    \caption{Plots concerning simulation scenario~III. The red color always refers to the Bayesian P-spline model and the green color to the proposed approach. In b), c), and d) the numbers 1 to 4 indicate the functional effect. a) Boxplots of the RMSE of the mean posterior predictions with respect to observation and to the true data generating function. b) Histograms of the posterior mean of the shrinkage coefficients. c) Boxplots of the RMSE of each estimated functional effect with respect to the true function. d) Boxplots of the RMSE of the estimated effects to the nearest parametric function. The RMSE in c) and d) is defined in terms of the integral from -1 to 1 instead of the mean of at observed values as in a).}
    \label{fig:sim3}
  \end{center}
\end{figure}

The plots in Figure~\ref{fig:sim3} summarize the main findings.
All plots are based on the estimated posterior mean.
Considering subplot a), we make the following observations.
First, the root mean square error (RMSE) with respect to the observed values is slightly smaller when using Bayesian P-splines.
Having said this, however, the RMSE with respect to the true values is smaller when employing our approach.
These observations indicate that the Bayesian P-spline slightly overfits.
The proposed prior seems to use the information supplied with the null space but is still able to diverge from it when it does not fit.
This interpretation is supported by the results obtained from the first two scenarios.

Breaking down the individual smooth effects, the histograms in Figure~\ref{fig:sim3}~b) show that the estimated shrinkage weight $\kappa$ is large for the splines with the correctly specified null space and small in the other case.
Thus, the prior is able to detect a misspecified null space and restrains the shrinkage.

This gets partly reflected in the RMSE (defined in terms of the integral from $-1$ to $1$ over the squared differences to the true function) of the individual smooth effects.
For the fourth smooth effect with the misspecified null space, the differences between the shrinkage prior and the P-spline solution are minor with respect to the RMSE.
The RMSE is the smallest and, in particular, considerably smaller than for the P-spline solution for the first and third smooth effect.
We find however almost no difference for the second smooth effect between both methods even though the shrinkage weight is in most replications close to one.
Together with the information provided in Figure~\ref{fig:sim3}~d), we deduce that the estimated function is within the null space, meaning it can be represented by a quadratic polynomial, but misrepresents the underlying true function.
The RMSE in this sub-figure shows the distance from the estimated function to the closest function from null space.
Again the RMSE is defined in terms of an integral.
Thus, the plots show that the functions estimated by the shrinkage approach are very close to a parametric solution from the null space for the first three functions.
For the misspecified null space, the estimated function is on average not close to a constant effect.

\subsection{Concluding remarks concerning the simulations}

To conclude, the proposed shrinkage prior is suited to take the additional information into account but can diverge from it when it is clearly misspecified.
This can help to avoid overfitting compared to the Bayesian P-spline.
However, an even better distinction between noise and signal is desirable, since this would lead to a more consistent estimation of extreme values of the shrinkage parameter $\kappa$, i.e., close to 0 or 1.
Values of this parameter far away from the extremes can be observed in scenario~I and scenario~II for some null spaces.

\section{Energy consumption in Germany over one day}\label{sec:energy}

Predicting energy consumption is an important aspect of keeping the energy grid stable.
Grid operators need to ensure that the amount of energy produced matches the amount of energy consumed.
However, failing to manage this successfully may cause small disturbances, such as fluctuations in the AC grid frequency, but also severe consequences such as blackouts.
To guarantee grid stability, the network operators have various options at their disposal.
Large and efficient power plants should provide the base load and more flexible power plants can be used to absorb rapid fluctuations in consumption.
Usually, energy produced by the flexible power plants is more expensive but large power plants, such as nuclear power plants, need more time to adapt their output.
Therefore, predicting the energy consumption that can be covered by the efficient power plants proves quite helpful.
Ideally, the long adaptation time of these power plants should be considered during estimation while still being able to flexibly detect quick changes in the base load.
The presented approach of smooth subspace shrinkage fulfills this requirement.
The subspace could be comprised of a trigonometric polynomial of low order that represents patterns that can be handled by efficient power plants.
However, a pure parametric estimation is not suitable, since a rapid increase in consumption, for example in the morning, must remain detectable.

In the following, we present an application to data from the German energy market.
The used data is freely available from SMARD\footnote{SMARD - Strommarktdaten der Bundesnetzagentur für Elektrizit\"at, Gas, Telekommunikation, Post und Eisenbahnen. The data can be downloaded at https://smard.de.}.
Suppose the quarter-hourly energy consumption in Germany over one day needs to be estimated.
The most naive approach might be a linear model based on a trigonometric polynomial of order $\Omega$, i.e.,
\[
  \operatorname{E}(y|x) = \beta_0 + \sum_{\omega=1}^\Omega \left[\beta_\omega\cos\left(\omega\frac{2\pi}{24}x\right) + \tilde\beta_\omega\sin\left(\omega\frac{2\pi}{24}x\right)\right]
\]
where $x \in [0, 24)$ denotes the hour of the day.
Another approach is the use of regression splines, as they can flexibly adapt to the data.
The latter may have the disadvantage that the spline solution diverges from the parametric solution even though the parametric solution is preferred based on theoretical considerations, and the spline solution fits the data only negligibly better.

We employ this example to showcase the applicability of the introduced prior hierarchy.
For that, we choose eight weekdays (only Mondays and Tuesdays) and eight weekend days from November~2018 and preprocess the data by rescaling and subtracting the daily mean.
We specify the parameters in our prior such that it shrinks towards the polynomial from above with $\Omega = 4$.
As explain above, we choose this order as we assume that efficient power plants can handle this pattern.

Since we employ only one covariate, we refrain from using the two level hierarchy for the scale parameter of the spline coefficients and thus fix $\xi$ to $0.001$.
Furthermore, the prior for $\tau$ is determined by setting the cutoff $c$ to two times the largest absolute second order derivative of the parametric solution and $\alpha = 0.05$.

The MCMC sampling runs for 12.000~iterations of which the first 2.000~are considered warmup.

\begin{figure}[hbt]
  \begin{center}
    \includegraphics[width=\textwidth]{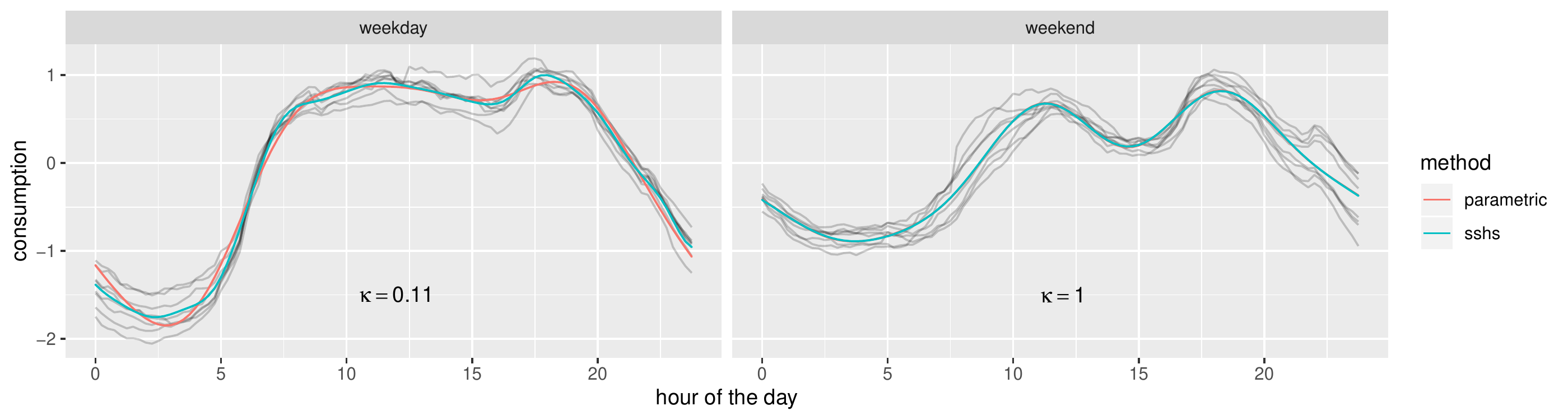}
    \caption{Plots of the total energy consumption in Germany for eight weekdays and eight weekend days in November~2018 (gray lines). Estimated outcome using the parametric and the introduced approach (sshs).}
    \label{fig:energy}
  \end{center}
\end{figure}

A plot of the data together with the estimated function is displayed in Figure~\ref{fig:energy}.
We observe that the proposed prior adopts to the data by shrinking the estimated effect to the parametric function for the weekend days and leaves it basically untouched for the other days.
This gets reflected in the shrinkage coefficient $\kappa$ that is almost one for the weekend, thus full shrinkage is observed.
Weekend days can be modeled with the chosen trigonometric polynomial while more flexibility is needed for weekdays.
Thus, our prior seems to trades off well between both situations.

\section{Discussion and concluding remarks}\label{sec:discussion}

In this paper, we presented a new prior hierarchy for smooth functional effects.
The novelty of the developed method is the ability to shrink towards a flexible user-defined null space while still enforcing smoothness for the estimate.
We showed that the prior is adaptive by shrinking small effects but leaving strong signals mainly untouched.

In simulation studies and an application, we demonstrated the performance of the shrinkage prior.
The empirical studies show that the prior works well in general, but under certain conditions a better separation between shrinkage and non-shrinkage is desirable.
To improve this, one could try to replace the half Cauchy distribution in the prior on $\lambda$.
A suitable candidate is the generalized centered half Cauchy distribution recently proposed by \citet{shinFunctionalHorseshoePriors2020}.
Using it implies, conditioned on $\tau \to \infty$, a beta prior on the shrinkage coefficient $\kappa$.
Consequently, the prior can be specified such that almost all probability weight is put on the extremes.
However, the proposed distribution does not include a scale parameter and, consequently, no global-local shrinkage is available.
The distribution would need to be enhanced to keep this feature.
The results of \citet{shinFunctionalHorseshoePriors2020} are promising and abandoning the global shrinkage parameter may be preferable.

We would also like to explore the option of specifying a joint prior on $(\lambda, \tau)$.
This might enable us to employ regularization of the curvature only if the shrinkage towards the subspace is weak. For strong shrinkage the smoothness is already guaranteed from the selected subspace.

Another focus of research could be the adaption of the proposed prior to more response distributions.
In a first step the prior could be adopted for the exponential family and then further extended to Bayesian distributional regression \citep{kleinBayesianStructuredAdditive2015}.
In this model class, all distributional parameters are related to the available covariate information via additive predictors.
Functional effects within the predictors could benefit from the proposed subspace shrinkage.

The proposed methodology could also be adopted in a penalized likelihood approach yielding a new type of shrinkage complementing LASSO-type methods.

\bibliography{references_sshs}

\appendix

\end{document}

%% file: 01_pkg_def.tex
\DeclareMathOperator{\ND}{N}
\DeclareMathOperator{\NDprec}{N_\text{prec}}
\DeclareMathOperator{\CD}{C}

\DeclareMathOperator{\InvGD}{IG}

\DeclareMathOperator{\Bfunc}{B}

\DeclareMathOperator{\rank}{rk}
\DeclareMathOperator{\EV}{E}

\DeclareBoldMathCommand\zerovec{0}
\DeclareBoldMathCommand\onevec{1}
\DeclareBoldMathCommand\betavec{\beta}
\DeclareBoldMathCommand\gammavec{\gamma}
\DeclareBoldMathCommand\epsvec{\varepsilon}
\DeclareBoldMathCommand\muvec{\mu}
\DeclareBoldMathCommand\thetavec{\theta}
\DeclareBoldMathCommand\etavec{\eta}
\DeclareBoldMathCommand\xvec{x}
\DeclareBoldMathCommand\zvec{z}
\DeclareBoldMathCommand\uvec{u}
\DeclareBoldMathCommand\svec{s}
\DeclareBoldMathCommand\wvec{w}
\DeclareBoldMathCommand\yvec{y}
\DeclareBoldMathCommand\Amat{A}
\DeclareBoldMathCommand\Bmat{B}
\DeclareBoldMathCommand\Fmat{F}
\DeclareBoldMathCommand\Imat{I}
\DeclareBoldMathCommand\Kmat{K}
\DeclareBoldMathCommand\Pmat{P}
\DeclareBoldMathCommand\Smat{S}
\DeclareBoldMathCommand\Qmat{Q}
\DeclareBoldMathCommand\Xmat{X}
\DeclareBoldMathCommand\Zmat{Z}